\documentclass[twocolumn]{aastex631}

\usepackage{aas_macros}
\usepackage{soul}
\usepackage{amsmath}
\usepackage{color,epsfig}

\begin{document}

\title[The impact of observation losses on IVS-R1/R4 VLBI sessions]{The impact of observation losses on IVS-R1/R4 VLBI sessions}

\author[0000-0001-9525-7981]{Tiege McCarthy}
\affiliation{University of Tasmania \\
School of Natural Sciences \\
Hobart, TAS 7005, Australia}

\author[0000-0002-7563-9488]{Lucia McCallum}
\affiliation{University of Tasmania \\
School of Natural Sciences \\
Hobart, TAS 7005, Australia}

\correspondingauthor{Tiege McCarthy}
\email{tiegem@utas.edu.au}

\begin{abstract}
Global VLBI observations, to measure Earth orientation and station positions, are organised into 24-hour sessions. Each session has a bespoke schedule created, optimised for the particular time period and the station network that is available during it. 
Due to various factors, whether it be station outages, sensitivity issues or source effects, not all scheduled observations are available, or of sufficient quality, to be included in the final geodetic analysis.
In this paper we derive statistics about the number of missing observations, as well as their effect on the expected precision of geodetic parameters such as station positions and Earth Orientation Parameters. 
We investigate the impact of observation loss on the weekly rapid turnaround IVS-R1 and IVS-R4 geodetic VLBI sessions over a decade period from $2014 - 2023$. 
Across our 1030 sessions we find on average 25.3\% of observations scheduled do not make it to analysis. 
This results in median performance losses, when compared to the scheduled versions, of 18.8\%, 19.2\%, 12.1/11.3\% and 28.7/22.9\% for UT1-UTC, 3D station position, X/Y nutation and x/y polar motion respectively. 
We find that the estimation of X/Y nutation is particularly robust to typical observation loss seen from these 24-hour sessions. 
Conversely, we see high-rates of critical degradation in performance (a doubling of the scheduled repeatability) for other geodetic parameters at observations losses of between 15--19\%, which is less than the median loss of 25.3\% that we find across this 10-year period.



\end{abstract}


\section{Introduction}\label{sec:intro}

One of the four space-geodetic techniques, geodetic very long baseline interferometry (VLBI) uses networks of radio telescopes to observe faint noise-like signals from extragalactic sources such as Quasars. These radio sources are sufficiently far away that they can be considered as fixed celestial reference points as observed from the Earth. The arrival time of the radio wave-front varies between telescopes depending on the geometry of the network and the motion of the Earth. This `group delay', the time delay measured between each pair of telescopes in the network, is referred to as an observation, and is the primary result of geodetic sessions. A typical geodetic session will produce many thousands of observations, which are compared against modeled delays with the residuals forming the basis of a Least-squares analysis that derives changes in the Earth orientation parameters (EOPs).

The premier VLBI sessions for the determination of EOPs are the two rapid turnaround IVS-R1 and IVS-R4 sessions, organised by the International VLBI service for geodesy and astrometry \citep[IVS;][]{nothnagel2017}. These series each consist of a weekly 24-hour session, with IVS-R1 starting on Mondays and IVS-R4 starting on Thursdays, with a global network and aim for a 15-day turnaround time for the geodetic results \citep{nothnagel2017}. These sessions have been in operation since 2002 and remain active today, with their networks evolving over time with a peak participation of 14 stations for both series reached in 2017 \citep{Thomas+24}. The next-generation VLBI Global Observing System (VGOS) promises to eventually replace these legacy dual band (S/X) sessions with broadband VLBI sessions \citep{niell2018}, however, challenges with roll-out of infrastructure and processing logistics means that the R1 and R4 sessions will remain relevant well into the future.

Schedules for the IVS-R1 and IVS-R4 sessions are typically distributed two weeks in advance. As each session is scheduled and optimised for a particular set of available stations and sources, station outages can cause significant impacts to the performance of EOP estimation. Outages that affect the optimisation of the schedule can often be unavoidable, with station issues only becoming apparent during the setup or observing of a session. In these cases it is not currently common to generate new schedules on-the-fly and distribute to all other stations, though systems to make this more feasible are being actively worked on \citep{Iles+18}. However, the impact of many outages is avoidable, and requires effective communication between the stations and the schedulers for when planned maintenance or extended outages will affect a stations ability to participate in upcoming sessions.

The motivation for this work is to understand the efficiency of the IVS network based on long-term statistics and highlight the impact of observation loss on the estimation of geodetic parameters. The observation of these sessions is still a very manual process, with many stations requiring input from operators for setup and monitoring. When final geodetic results are impacted by observation loss from operator mistakes, insufficient performance monitoring or failures of communication within the network, more investment into the development of new ways to operate the network are warranted (e.g. feedback loops, automation of station setup).

This paper analyses the past decade of IVS-R1 and IVS-R4 sessions in order to determine the fraction of scheduled observations that are lost prior to analysis. We then investigate the impacts of these observation losses on the final EOPs by comparing simulated performance of each scheduled session, to the simulated performance of the same sessions with the observed data loss. This will allow us to determine how much performance is being degraded due to misseing data and which EOPs are most adversely affected. Additionally, it will also provide a scale factor, based on the 941 sessions analysed, that can be applied to future simulation studies to take into account the expected performance impacts experienced by real-world observing programs.

\begin{table*}

\caption{Median relative performance values over the entire dataset (2014 through 2023). This dataset contains a total of 1030 sessions (515 R1s and 515 R4s). For estimated parameters (EOPs and 3D station coordinates), these performance values are the inverse of the repeatability values as determined by simulating observations usable in analysis over the theoretical repeatability (i.e. the fraction of theoretical performance achieved).}
\begin{center}
\begin{tabular}{ccccccccc}
\hline
\multicolumn{1}{c}{\bf Series} & {\bf Sessions} & {\bf no. obs.} & {\bf UT1-UTC} & {\bf X pol. } & {\bf Y pol.} & {\bf X nut.} & {\bf Y nut. } & {\bf 3D coord}      \\ \hline
{\bf R1} &     515 &   0.751   &  0.824  &  0.729  & 0.783 &  0.882  &  0.888 & 0.835     \\
{\bf R4}  &    515 &   0.743   &  0.799  &  0.691  & 0.743 &  0.878  &  0.886 & 0.726 \\
{\bf Total} &  1030 &   0.747   &  0.812  &  0.713  & 0.771 &  0.879  &  0.887 & 0.808  \\ 
\hline
\end{tabular}
\end{center} \label{table:full_dataset_table}
\end{table*}

\begin{figure}
\begin{center}    
\epsfig{figure=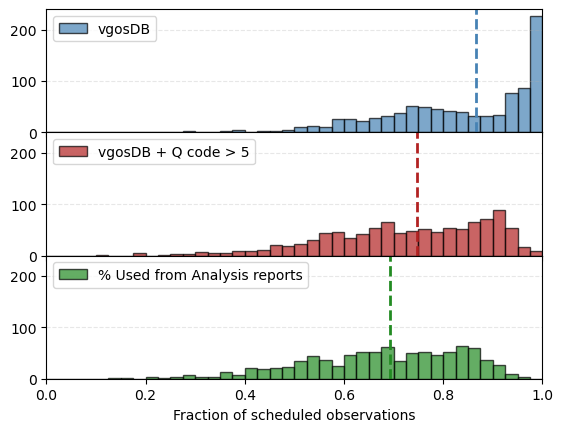,scale=0.60}
\caption{{\it Top panel:} Distribution (number of sessions) for the number of observations contained within the vgosDB of a session versus scheduled. {\it Middle panel:} Same as above, with the additional filter of the observations requiring a quality code $\ge5$. {\it Lower panel:} Distribution  (number of sessions) of number of observations used versus scheduled in R1 and R4 IVS analysis reports over the same 10 year period. The vertical dashed lines represent the median values for each distribution.}
\label{fig:obs_hists}
\end{center}
\end{figure}

\section{Methodology} \label{sec:method}

\subsection{Sample selection} \label{sec:sample}

We utilise the simulator within VieSched++  \citep{Schartner+19} to compare the performance of the theoretical best-case scheduled scenario for a given session, all observations present in the database, versus the `observed' case, only the observations that are usable in analysis (see Section \ref{sec:sims} for specific details). A total of 1030 rapid 24 hour VLBI sessions (R1 and R4 series), over a one decade time period (2014 -- 2023) were considered in this analysis. Both the schedule files, and vgosDB geodetic databases were downloaded from CDDIS for these 1030 VLBI sessions. The vgosDB geodetic databases for a session contains all observations that were correlated for a particular session (i.e. any observations missed due to station outages or missing data will not be present). From these vgosDB databases, we generated lists of observations that can be used to limit which observations are considered during simulation. In order to determine the most realistic sample, we considered two different criteria when selecting observations for our `observed case'. The first, was to include all observations present in the vgosDB database. This would capture the effect of any data losses due to station outages or critical recording errors where the observations were discarded during correlation, however, the effects of non-detections and poor quality observations due to station sensitivity issues would be ignored. This selection criterion resulted in a median observed over scheduled observation ratio of 0.864. The alternate case was to further filter the sample, including only observations that had an associated `quality code' (a measure of the fringe quality) of $\ge5$, essentially removing any observations that were of too poor quality to be included in the analysis. This more restrictive selection criteria resulted in a median observed over scheduled observation ratio of 0.747. In order to determine whether our more restricted dataset was realistic, we extracted the used/scheduled observation fraction from the IVS analysis reports (referred to as `performance' within these reports) of sessions over this same one decade time period and determined the median fraction of scheduled observations that make it through to analysis is 0.693. While this is lower than the median value of our restricted sample, this is to be expected, as occasionally data is excluded  by analysts for other reasons, such as on the baselines between co-located stations (e.g.  the baseline between the WETTZELL and WETTZ13N telescopes at the Wettzell Geodetic Observatory). Figure~\ref{fig:obs_hists} compares the distributions for these 3 different datasets, and further supports the usage of the more restrictive dataset for our investigation into the impact of data losses. All data we utilise in this investigation (vgosDB, schedule files and analysis reports) is publicly available on CDDIS or via the IVS master schedule.

Each session was run through the simulation process twice, first using every observation defined by the schedule file, and then a second time restricted to only those present in the vgosDB with quality codes $\ge5$. The results of these simulations are repeatability values for each EOP, in the scheduled and the observed case for each of the 1030 sessions. Simulated, rather than real values for the `observed' case are used in order to compare values in the most equivalent way possible.   It should be noted, observations that are unique to the `observed' case, such as the ad-hoc addition of tag along stations, are removed as they do not have equivalents from the `scheduled' version.

Each VLBI session differs greatly from one to the other (depending on the available network), with the number of observations sometimes varying by a factor of 3 within the same series \citep{Thomas+24}. As we want to directly compare the theoretical performance of sessions to their realised counterparts, we instead consider the relative performance between the `ideal' and `observed' values of each session. As most parameters are repeatability values (i.e. higher values indicate worse performance) we consider the inverse of the observed over ideal fraction. The only parameter that is an exception to this is number of observations, where we consider the number of observations `observed' over the number scheduled. For example, if the observed case of a session had 25\% fewer observations and twice the repeatability for a particular EOP, the relative performance of observations would be 0.75, and of the EOP would be 0.5. 

\subsection{Simulations and analysis}  \label{sec:sims}

The VieSched++ simulator utilises Monte Carlo simulations \citep{Pany2011} before performing a least-squares based geodetic VLBI analysis to esimate the expected precision of EOPs and station positions \citep[e.g.][]{Schuh2013}.
A white noise of 25~ps per observation was used to simulate measurement noise (corresponding to 17.68~ps per station, a conservative estimate for S/X stations).
Clock drifts were simulated using an integrated random walk process with an Allan standard deviation of $1 \times 10^{-14}{s}$ after 50 minutes. Effects of tropospheric delays were simulated following \citet{Nilsson2010} using an average tropospheric turbulence factor $C_n$ of $1.8\times 10^{-7}{m^{-1/3}}$. 

Within the least-squares analysis, each parameter is estimated as offsets beside piece-wise linear clock drifts (with an estimation interval of 60 minutes) per station (except for one station set as the reference), zenith wet delay (interval of 30 mins) and atmospheric gradients (interval of 180 mins). Station coordinates and EOPs were estimated once per 24 hour session with source coordinates remaining fixed. All stations were utilised in datum realisation.

For both cases of each session (the scheduled version and the `observed' version) 1000 simulation and analysis runs were generated to determine accurate repeatability values for each EOP together with the reported mean formal errors from the least-squares-based variance-covariance matrix.

\begin{figure*}[h]
\begin{center}    
\epsfig{figure=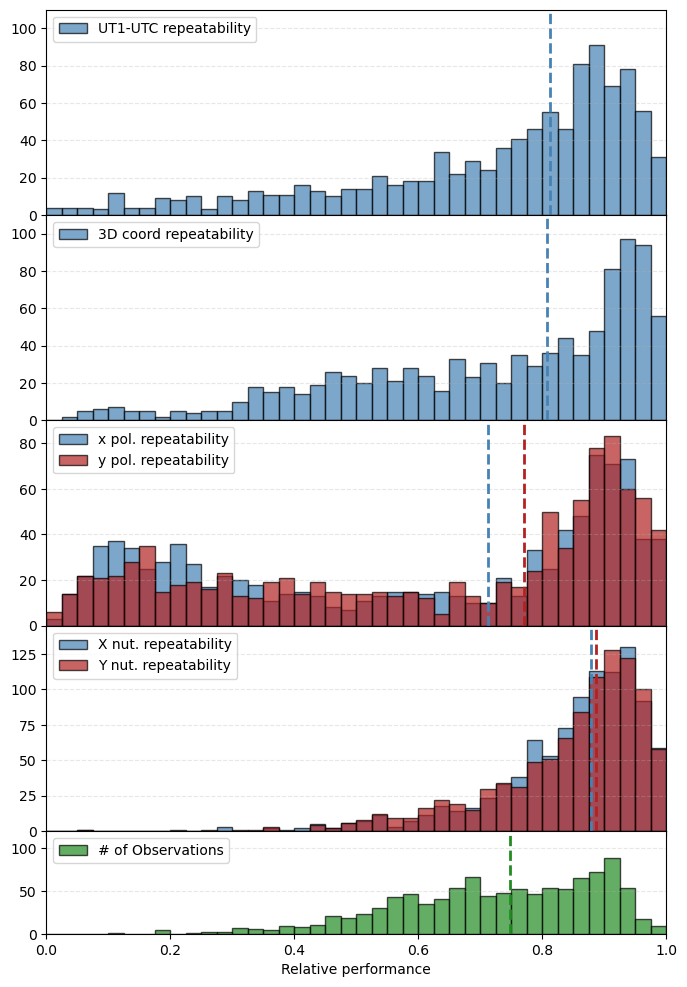,scale=0.80}
\caption{Histograms (number of sessions) of parameter performance between the session as observed versus the theoretical perfect session. For example, a value of 0.5 represents the observed session having twice the repeatability of the ideal scheduled case case. Vertical dashed line represents the median value of each distribution. }
\label{fig:all_hists}
\end{center}
\end{figure*}

\section{Results} \label{sec:results}

The relative performance values derived from our simulations are presented in Table~\ref{table:full_dataset_table}, which contains the median values across the full 10 year dataset, broken down by R1 and R4 session, in addition to the combined total values. Annual median values are provided in Table~A1 of the appendix.

\subsection{Combined 10 year dataset}\label{sec:result_10yr}

When considering the full dataset (see Figure \ref{fig:all_hists} for the distributions), we see a median observation performance of 0.747, indicating that 25.3\% of scheduled observations  are either not observed, or alternatively, not to an appropriate standard for use in analysis. This 25.3\% loss in observations results in median parameter estimation performance losses of 18.8\% for UT1-UTC, 19.2\% for 3D station position, 12.1/11.3\% for X/Y nutation and 28.7/22.\% for x/y polar motion.

\begin{figure*}[ht]
\begin{center}    
\epsfig{figure=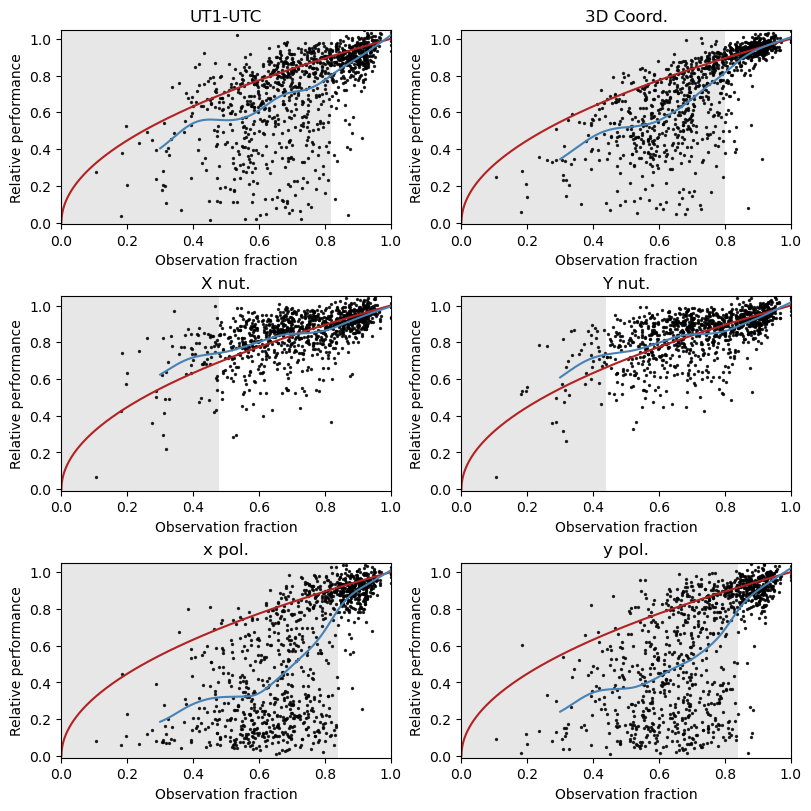,scale=0.70}
\caption{Scatter plot for each parameter of relative performance versus fraction of observations that make it to analysis. Red lines represent the square root relationship ($f(x) = \sqrt{x}$) that is expected from a sample of independent, equally weighted data points. The blue smoothing spline displays the broad trend of the data for ease of comparison. Shaded areas represent the minimum levels of observation loss where 10\% or more sessions have double the scheduled repeatability (relative performance $<0.5$).}
\label{fig:all_scatter}
\end{center}
\end{figure*}

Figure \ref{fig:all_scatter} shows scatter plots of the performance scaling versus the fraction of observations present for analysis. For most parameters we see a broadly linear scaling, with relatively low variance, between the performance of parameter estimation and missing observations relatively tight variance at low levels of observation loss ($>0.85$ observation fraction).  This relationship continues until reaching a critical level of observation loss where variance significantly increases, and sessions becoming increasingly likely to suffer from catastrophic performance degradations. X/Y nutation estimation appears to be much more robust to observation loss compared to the other estimated parameters, with far fewer sessions critically impacted. The details of these plots, and performance breakdowns despite relatively high observation fractions are discussed further in Section \ref{sec:param_estimation}.

\begin{figure}
\begin{center}    
\epsfig{figure=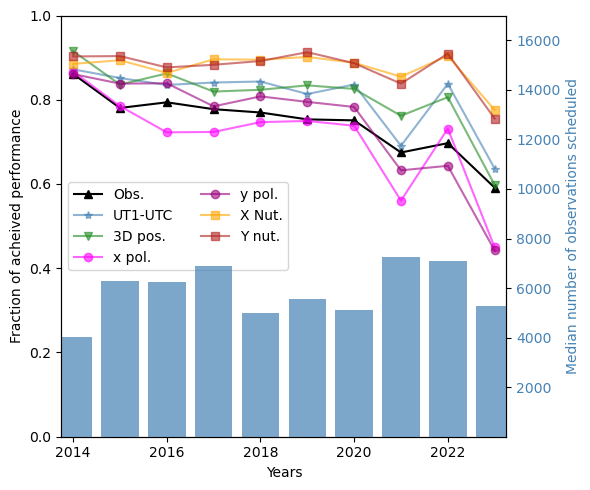,scale=0.60}
\caption{Annual median values of relative performance for all considered parameters. The blue columns represent the annual median number of observations scheduled per session.}
\label{fig:performance_by_year}
\end{center}
\end{figure}

\subsection{Annual performance}

In addition to the full 10 year dataset, we can look at the annual median values to investigate how observation loss has impacted our performance over the past decade. 
Figure \ref{fig:performance_by_year} shows that, despite a relatively consistent network size (and subsequent number of scheduled observations), there is a general downward trend in how many observations are observed versus scheduled across this ten year period. We can see from this figure that most of the estimated parameters broadly follow the same behaviour as seen in the observation fraction, with the level of impact dependent on which particular parameter is being considered. Notably, X and Y nutation values appear to show less negative impact from observation loss, whereas, estimates of polar motion appear to be the most heavily impacted. This large impact in the polar motion parameters can be particularly seen in the 2023 R4 subset where a median observation percentage of 56.8\% resulted in median relative performance values of 26.9 and 25.2\% for the X and Y polar motion parameters respectively.

It is important to re-iterate, these are ratios relative to scheduled performance, meaning the downtrend over time does not necessarily indicate our estimation of these parameters is getting worse over time \citep[fortunately the opposite is actually the case; see][]{Thomas+24}. Instead it is indicating that we are extracting less of the absolute scheduled performance potential out of the average session as the years progress.

\section{Discussion and analysis}\label{sec12}


\subsection{Observation losses in IVS-R1 and IVS-R4 sessions}


 The consideration of what observations to include in our simulations outlined in Section \ref{sec:sample} allows for some insight into the relative fractions of observation losses that are due to missing data  versus those that are non-detections or of insufficient quality for analysis. We can see that the median session has 13.4\% of observations never making it past correlation, with the majority of these because of station non-participation (while still included in the schedule), or significant setup/recording issues resulting in no valid data. When further restricting the sample to only good quality observations (detections with a Q-code $\ge5$) we see an additional 11.9\% loss of scheduled observations, bringing the total observation loss up to 25.3\%, over a quarter of scheduled observations. Considering the median network size of 10 stations across our dataset, and assuming equal observation participation across all stations, this corresponds to a median per-station efficiency of $\sim87\%$, similar to the 90\% reported by \citet{Thomas+24} across the period of 2002--2020. These lost observations can be caused by a number of issues, including local unwanted electromagnetic emissions at stations, sensitivity degradations with hardware or incorrect assumptions during scheduling (either for station sensitivity or source brightness/structure). In order to determine with more granularity the specific causes for observation loss requires a much deeper investigation into the correlation and analysis reports that are produced during the routine processing and analysis of these sessions and is outside the scope of this investigation.

\subsection{Effect of observation loss on geodetic parameter estimation} \label{sec:param_estimation}


 The scatter plots of Figure \ref{fig:all_scatter} allow us to visualise the performance scaling across the varying levels of observations losses the IVS-R1 and IVS-R4 sessions experience. These scatter plots have been overlaid with a smoothing spline to capture general trend of the data points \citep[blue line; ][]{Woltring86} and  square root relationship (red line). As mentioned in Section \ref{sec:result_10yr}, this square root relationship represents the performance scaling expected from observation losses, if all observations are uncorrelated, independent and equally weighted. All 6 parameters show relatively smooth performance fall-off at minor levels of observation losses (observation fraction $> 0.85$), with relatively low variance. As the observation fraction further decreases, we start to see dramatically increased performance variance in all parameters with the notable exception of X/Y nutation, which remains consistent with the square root relationship scaling. In comparison, the other parameters display much more significant deviations from the square root scaling, with x/y polar motion the most pronounced. A potential explanation is that the estimation of these parameters is more dependent on specific network geometry, for example, north-south baselines are particularly sensitive to x/y polar motion, and east-west baselines sensitive to UT1-UTC \citep{Dermanis&78, Schartner+20} and loss of observations with these orientations will have a larger impact on the parameter estimation.


 In addition to the general scaling of performance versus observation loss, we also see that performance can show significant degradations despite the observed/scheduled ratio being high for a particular session.  This is likely due to these sessions suffering the loss of `higher value' observations, such as from baselines with higher-impact geometry (e.g. east-west baselines for UT1-UTC) or from station outages from remote stations that may contribute relatively few observations but dramatically increase the total extent of the network. These major performance breakdowns are of most concern, as they indicate that the schedule for the session has been significantly compromised. From Figure \ref{fig:all_scatter} we can see that some sessions with observation losses of as little as 10\% may experience this severe reduction in parameter estimation performance.


If we define a halving of performance (i.e. a doubling of the scheduled repeatability value) as a critical performance breakdown, we can determine the total fraction of sessions at or below this achieved performance level.  X/Y nutation only suffers critical degradations in $\sim2\%$ of sessions, UT1-UTC and 3D station position estimation are the next best performing, with 17\% and 19\% of sessions affected respectively, and on the opposite end of the scale x/y polar motion estimation is critically affected in 39\% and 36\% of sessions respectively. 
Additionally, we bin the data displayed in Figure \ref{fig:all_scatter} into 0.02 observation fraction bins and determine at which point $>10\%$ of sessions within a bin suffer a critical breakdown. 
X and Y nutation again perform the best, with the centre observation fractions of the first bins fitting this criteria at 0.49 and 0.45 (i.e. 51\% and 55\% observation losses). On the other hand, the other four parameters reached this 10\% degradation cutoff at 19\% observation loss for 3D station position, 17\% observation loss for UT1-UTC estimation, with the X and Y polar motion parameters performing the worst, reaching the breakpoint at 15\% observation loss. This means that at observation losses as low as 15\%, one in ten sessions is failing to measure polar motion to within a factor of 2 of the predicted precision.
The lack of robustness in the polar motion parameters may be the result of it's sensitivity to north-south baselines \citep{Schartner+20}. The R1 and R4 networks have relatively few southern-hemisphere stations and observation loss from these stations will significantly reduce the number of north-south baselines, resulting in a disproportionate impact on the polar motion estimation of an affected session.
These break points (for all parameters apart from X/Y nutation) are significantly higher than the median observation fraction we see across the 10 year dataset, indicating that a typical R1 or R4 sessions is at risk of performing significantly worse than expected when experiencing typical levels of observation loss.

\begin{table}
\begin{center}
\caption{Results from Kolmogorov-Smirnov and Mann-Whitney U tests on distribution differences between results from R1 and R4 data subsets. The $D_{KS}$ statistic represents the maximum difference between the two distributions. $p_{KS}$ and $p_{MWU}$ are the p-values for the Kolmogorov-Smirnov and Mann-Whitney U tests respectively.}
\begin{tabular}{lccc}
\hline
\multicolumn{1}{c}{\bf Parameter} & $D_{KS}$ & {\bf $p_{KS}$} & {\bf $p_{MWU}$}  \\ \hline
 nObs         & 0.076 & 0.104    & 0.264   \\
 UT1 -- UTC   & 0.109 & 0.005    & 0.008   \\
 3D Coord.    & 0.272 & $<0.001$ & $<0.001$   \\
 X nut.       & 0.039 & 0.833    &  0.935    \\
 Y nut.       & 0.054 & 0.432    &  0.625  \\
 X pol.       & 0.155 & $<0.001$ &  0.001  \\
 Y pol.       & 0.128 & $<0.001$ &  0.004  \\
\hline \label{tab:tests}
\end{tabular}
\end{center}
\end{table}

\subsection{Comparison of R1 and R4 sessions} \label{sec:r1_vs_r4}

The median performance values between the R1 and R4 subset of the 10 year dataset are within 5\% for all parameters except 3D station position which shows a more significant 16\% difference (values in Table \ref{table:full_dataset_table}). Despite this, we see differences in the overall distribution of performance values between the two series in some parameters. We utilise both the Mann-Whitney U test \citep[MWU; ][]{mann&47}, and Kolmogorov-Smirnov test \citep[KS; ][]{smirnov39} to compare the distributions between the R1 and R4 subsets of each parameter. The null hypothesis of both of these tests are that both distributions are identical with the MWU test sensitive to changes in median, and KS test sensitive to differences in distribution shape.
A p-value of $< 0.05$ represents a rejection of the null-hypothesis, indicating the two distributions are statistically different. Across both tests the distributions for number of observations, and X/Y nutation are considered identical. However, even with similar observation distributions we see p-values $< 0.05$ for both the KS and MWU tests for UT1-UTC, mean 3D station coordinates, and x/y polar motion indicating that these parameters have differing performance scaling between the IVS R1 and R4 observing series. Qualitatively, when looking at the performance distributions (see Figure \ref{fig:r1vsr4}) we see that for 3D station position and x/y polar motion, the R4 distributions have a secondary peak at lower values of relative performance (0.1 for x/y polar motion, and $\sim0.5$ for station positions).

Across our 10 year dataset, the R4 network generally had sparser networks in the southern hemisphere and eastern Asia \citep{Thomas+24}. For example, across our dataset 5 stations in east Asia regularly participate in the R1 network (SEJONG, ISHIOKA, TSUKUB32, KASHIM34, SESHAN25), in stark contrast to the single east Asian station regularly participating in R4 sessions (ISHIOKA). This lack of redundancy for the R4 sessions means that a relatively small amount of observation loss from this station can significantly reduce the total network extent, and disproportionately affect the performance of the session.



\subsection{Simulation scale factor} \label{sec:sim_scaling}

Simulations are heavily used for the determination and optimisation of new geodetic VLBI observing programs \citep[e.g.][]{Schartner+20, Bohm+22, Dhar+23, Wolf&23,Laha+24, Schunck+24, McCarthy+25}. These simulations assume a 100\% success rate for observations which is useful for optimisation but doesn't necessarily reflect reality. The 10 year median performance values in Table \ref{table:full_dataset_table} can be used as a base point for scaling of simulation values to account for realistic observation loss. However, it should be noted that these median performance values are based on the IVS R1 and R4 networks, which are global networks conducting regular 24-hour S/X sessions. Median values for VGOS sessions may be different due to the different networks, and potentially different levels of observation loss as the technique is being developed. It is also likely that regional networks, or those with less total extent, will see different performance scaling with observation losses. 

\section{Conclusion}\label{sec13}

Across our 10 year dataset, we identify a median session observation loss of 25.3\% for IVS R1/R4 sessions, with the annual median trending downward across this period, implying a per-station efficiency of $\sim87\%$. This level of observation loss corresponds to a median performance reduction of 18.8\%, 19.2\%, 12.1/11.3\% and 28.7/22.9\% across the UT1-UTC, 3D station positions, X/Y nutation and x/y polar motion parameters respectively. All parameters except for X/Y nutation show significant percentages of sessions being affected by critical breakdowns in performance (a doubling of repeatability values), with x/y polar motion the worst affected at 39\% and 36\% of sessions. These critical breakdowns begin to affect more than 10\% of sessions at levels of observation loss between 15--19\%, dependent on the parameter.


When considering the distribution for the R1 and R4 sessions separately, we see differences in performance scaling across all parameters except X and Y nutation. Differences between the two series are particularly pronounced for the 3D station position and X  polar motion parameters.

As a result of this work, we recommend the application of a scale factor to determine more realistic performance values (accounting for typical data losses) when simulations are used to determine expected precision of estimated parameters. In particular, this is useful when absolute values are being presented.


An interesting area of further investigation would be determining how performance can be recovered through re-scheduling in cases where station non-participation is known prior to the session start time. This re-scheduling may prevent cases where the presence of a non-participating station was critical to the schedule structure and has a disproportionate impact. However, short notice re-scheduling does present some logistical issues, with special care needing to be taken to ensure all stations are on a consistent version of the session schedule prior to observations starting.

Reduction of this high rate of data loss must be a priority for these sessions moving forward to effectively minimise the impact seen on final geodetic results. Improvements can be made to various aspects of the geodetic VLBI pipeline in order to achieve this. Increased proficiency in station operation is one such avenue, with development of tools for performance monitoring and use of automation for consistent station setups. Another avenue is better communication of common or re-occurring problems seen by correlators or analysts back to the operations teams at stations. Observation planning and scheduling is also an important tool for minimising the impact of data loss, through the choice of networks and observing strategies that are more robust to typical levels of observation loss.

\section*{Acknowledgments}
We thank the reviewers for their insightful comments which have helped improve the manuscript. The AuScope VLBI project is managed by the University of Tasmania, contracted through Geoscience Australia.
This work was supported by the Australian Research Council (DE180100245). 
This research has made use of NASA’s Astrophysics Data System Abstract Service. 

\section*{Author Contributions}

TM and LM developed the study conception and design. TM developed the scripts required to generate results. TM and LM worked on interpretation of the results. TM completed the first draft of the manuscript. LM provided comments, suggestions and additional text to develop the final manuscript. Both authors read and approved the final manuscript.

\section*{Data Availability}

The schedule files, vgosDB geodetic databases and analysis reports used for this investigation are all publicly available through CDDIS (or other IVS datacentres) or the IVS master schedule (ivscc.gsfc.nasa.gov). Complete simulation results are available from the authors upon reasonable request (tiegem@utas.edu.au).

\bibliography{references}
\bibliographystyle{aasjournal_edit}

\appendix

\counterwithin{figure}{section}
\counterwithin{table}{section}
\renewcommand\thefigure{\thesection\arabic{figure}}
\renewcommand\thetable{\thesection\arabic{table}}

\section{Appendix 1}

\begin{table*}[h!]
\begin{center}  \label{table:annual_medians}
\caption{Median relative performance values over the entire dataset (2014 through 2023), presented as both annual values, and as a combined dataset. This dataset contains a total of 1030 sessions (515 R1s and 515 R4s). For estimated parameters (EOPs and 3D station coordinates), these performance values are the inverse of the repeatability values as determined by simulating observations usable in analysis over the theoretical repeatability (i.e. the fraction of theoretical performance achieved). Please see the supplementary information section if you would like access to the full dataset with individual session information.}
\begin{tabular}{lccccccccc}
\hline
\multicolumn{1}{c}{\bf Year} & \multicolumn{1}{c}{\bf Series} & {\bf Sessions} & {\bf no. obs.} & {\bf UT1-UTC} & {\bf X pol. } & {\bf Y pol.} & {\bf X nut.} & {\bf Y nut. } & {\bf 3D coord}      \\ \hline
{\bf 2014}     &  \bf R1  & 49  & 0.859  &   0.870  &  0.871  & 0.854 &   0.885 &  0.903 & 0.923    \\
              &  \bf R4  &  50  & 0.865  &   0.874  &  0.860  & 0.874 &   0.889 &  0.910 & 0.912    \\
             &  \bf Total & 99 &  0.861   &  0.872  &  0.867  & 0.861 &   0.886 &  0.903 & 0.915     \\
{\bf 2015}    &  \bf R1  &  52 &  0.768  &   0.832 &  0.796  & 0.853 &   0.884 &  0.903 & 0.860 \\
              & \bf  R4  &  52 &  0.797  &   0.866  &  0.723  & 0.813 &   0.907 &  0.904 & 0.804     \\
             & \bf Total &  104 &  0.790  &   0.852  &  0.785  & 0.838 &   0.894 &  0.904 & 0.836     \\
{\bf 2016}    & \bf R1  &  52  &   0.764  &   0.865  &  0.785  & 0.871 &    0.879 &  0.890 & 0.878     \\
             & \bf  R4  &  52  &    0.798   &  0.775  &  0.672  & 0.816 &   0.852 &  0.867 & 0.817     \\
             & \bf Total &  104 &   0.794   &  0.835  &  0.723  & 0.839 &   0.864 &  0.877 & 0.862     \\
{\bf 2017}      & \bf R1 & 49 &  0.793 &  0.856  &  0.805   & 0.810 &  0.915 &   0.908 & 0.887 \\
               & \bf R4 & 49 &   0.765 &   0.795  &  0.652  & 0.651 &   0.875 &  0.864 & 0.760     \\
            & \bf Total & 98 &   0.778 &   0.841  &  0.724  & 0.784 &   0.896 &  0.882 & 0.819     \\
{\bf 2018}    & \bf R1 & 52 &  0.800  &   0.872  &  0.849  & 0.861 &   0.897 &  0.903 & 0.852     \\
              & \bf R4 & 52 &  0.704   &  0.806  &  0.676  & 0.764 &   0.884 &  0.886 & 0.678     \\
          & \bf Total & 104 &  0.770   &  0.843  &  0.747  & 0.808 &   0.895 &  0.892 & 0.824     \\
{\bf 2019}  &  \bf R1  & 53  &  0.722  &   0.814  &  0.711  & 0.693 &  0.900 &   0.896 & 0.834 \\
            &  \bf R4  & 52  &  0.821  &   0.825  &  0.772  & 0.852 &   0.920 &  0.928 & 0.834     \\
           & \bf Total & 105 &  0.753  &   0.814  &  0.750  & 0.795 &   0.902 &  0.913 & 0.834   \\
{\bf 2020} &  \bf R1 & 52 &  0.765  &  0.846  &  0.746  & 0.796 &   0.894 &  0.891 & 0.848     \\
            & \bf R4 & 53 &  0.703  &  0.809  &  0.664  & 0.660 &   0.881 &  0.879 & 0.721     \\
         & \bf Total & 97 &  0.751  &  0.836  &  0.738  & 0.783 &   0.888 &  0.887 & 0.826     \\
{\bf 2021} & \bf  R1 & 52 &  0.677   &   0.690  &  0.543   & 0.653 &  0.850 &  0.849 & 0.773 \\
           & \bf R4 & 52  &  0.623   &   0.710  &  0.622  & 0.596 &   0.858 &  0.836 & 0.653     \\
        & \bf Total & 104  &  0.675   &   0.691  &  0.560  & 0.632 &   0.855 &  0.838 & 0.762     \\
{\bf 2022}& \bf  R1 & 52 &  0.713   &    0.847  &  0.687  & 0.696 &  0.888 &  0.897 & 0.817     \\
           & \bf R4 & 52 &  0.682   &    0.823  &  0.748  & 0.565 &  0.910 &  0.914 & 0.698     \\
        & \bf Total & 104 &  0.697   &   0.837  &  0.731  & 0.643 &   0.904 &  0.910 & 0.806     \\
{\bf 2023} & \bf  R1 & 52 & 0.656  &   0.686  &  0.591   & 0.529 &  0.786 &   0.759 &  0.696 \\
           & \bf R4 & 51 &  0.568 &    0.542  &  0.269  & 0.252 &   0.766 &  0.743  & 0.512     \\
        & \bf Total & 103 & 0.591 &    0.637  &  0.450  & 0.442 &   0.776 &  0.755  & 0.598     \\
\hline
{ \bf Full} & {\bf R1} &     515 &   0.751   &  0.824  &  0.729  & 0.783 &  0.882  &  0.888 & 0.835     \\
            & {\bf R4}  &    515 &   0.743   &  0.799  &  0.691  & 0.743 &  0.878  &  0.886 & 0.726 \\
            & {\bf Total} &  1030 &   0.747   &  0.812  &  0.713  & 0.771 &  0.879  &  0.887 & 0.808  \\ 
\hline
\end{tabular}
\end{center} 
\end{table*}

\begin{figure*}[ht]
\begin{center}    
\epsfig{figure=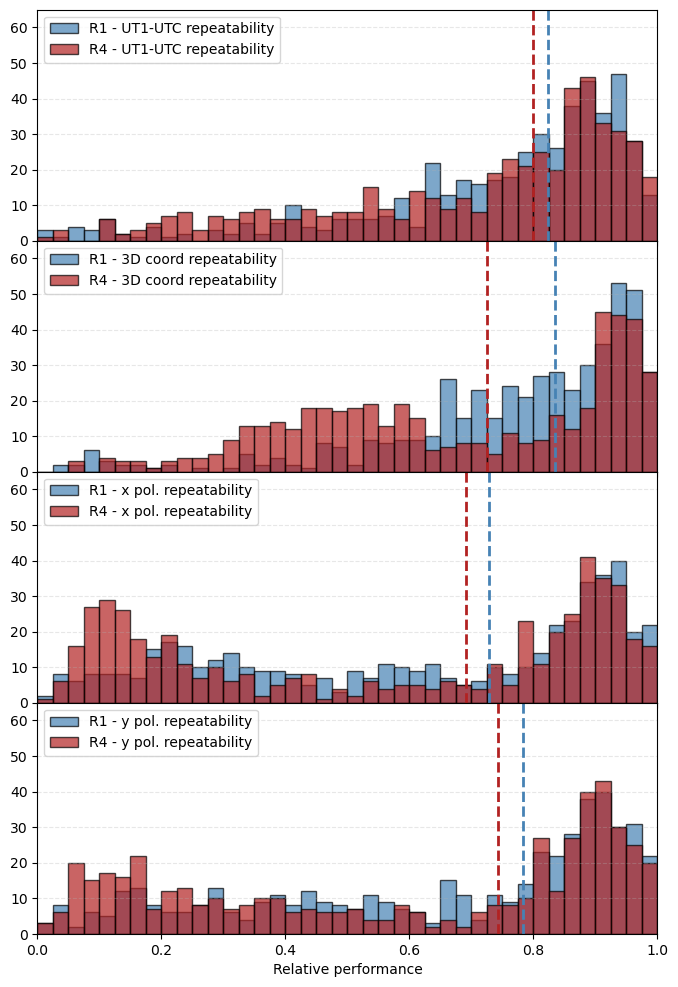,scale=0.80}
\caption{Histograms (number of sessions) of parameter performance comparing R1 and R4 sessions over the 10 year period. Vertical dashed line represents the median value of each distribution. Only parameters with distributions deemed significantly different (based on the Kolmogorov-Smirnov tests discussed in Section \ref{sec:r1_vs_r4}) have been included here for comparison.}
\label{fig:r1vsr4}
\end{center}
\end{figure*}



\end{document}